\documentclass[twocolumn,prd,nofootinbib
,floatfix ,showpacs,superscriptaddress]{revtex4}
\usepackage{graphicx}

\pdfoutput=1

\begin{document}

\title{Probing the leptonic Dirac CP-violating phase in neutrino oscillation experiments}

\author{Tommy Ohlsson}
\email{tohlsson@kth.se}
\affiliation{Department of Theoretical Physics, School of
Engineering Sciences, KTH Royal Institute of Technology, AlbaNova
University Center, 106 91 Stockholm, Sweden}

\author{He Zhang}
\email{hzhang@mpi-hd.mpg.de}
\affiliation{Max-Planck-Institut f\"{u}r Kernphysik, Saupfercheckweg
1, 69117 Heidelberg, Germany}

\author{Shun Zhou}
\email{shunzhou@kth.se} \affiliation{Department of Theoretical
Physics, School of Engineering Sciences, KTH Royal Institute of
Technology, AlbaNova University Center, 106 91 Stockholm, Sweden}

\begin{abstract}
The discovery of leptonic CP violation is one of the primary goals
of next-generation neutrino oscillation experiments, which is
feasible due to the recent measurement of a relatively large
leptonic mixing angle $\theta^{}_{13}$. We suggest two new working
observables $\Delta A^{\rm m}_{\alpha \beta} \equiv \max[A^{\rm
CP}_{\alpha \beta}(\delta)] - \min[A^{\rm CP}_{\alpha
\beta}(\delta)]$ and $\Delta A^{\rm CP}_{\alpha \beta}(\delta)
\equiv A^{\rm CP}_{\alpha \beta}(\delta) - A^{\rm CP}_{\alpha
\beta}(0)$ to describe the CP-violating effects in long-baseline and
atmospheric neutrino oscillation experiments. The former signifies
the experimental sensitivity to the leptonic Dirac CP-violating
phase $\delta$ and can be used to optimize the experimental setup,
while the latter measures the intrinsic leptonic CP violation and
can be used to extract $\delta$ directly from the experimental
observations. Both analytical and numerical analyses are carried out
to illustrate their main features. It turns out that an intense
neutrino beam with sub-GeV energies and a baseline of a few
$100~{\rm km}$ may serve as an optimal experimental setup for
probing leptonic CP violation.
\end{abstract}


\maketitle

\section{Introduction}

The phenomenon of neutrino oscillations in vacuum and matter can be
described by six fundamental parameters: three lepton flavor mixing
angles $\{\theta^{}_{12}, \theta^{}_{13}, \theta^{}_{23}\}$, two
neutrino mass-squared differences $\{\Delta m^2_{21}, \Delta
m^2_{31}\}$, and one Dirac-type CP-violating phase $\delta$. Due to
a number of elegant neutrino oscillation experiments in the past
decades, both $\{\theta^{}_{12}, \Delta m^2_{21}\}$ and
$\{\theta^{}_{23},|\Delta m^2_{31}|\}$ have been measured with
reasonably good accuracy \cite{Beringer:1900zz}. Until recently, the
smallest leptonic mixing angle $\theta^{}_{13}$ has been found to be
relatively large in the Daya Bay \cite{An:2012eh} and RENO
\cite{Ahn:2012nd} reactor neutrino experiments. This great discovery
enhances the capability of the next-generation experiments to pin
down the neutrino mass hierarchy (i.e., the sign of $\Delta
m^2_{31}$) and eventually to determine the leptonic Dirac
CP-violating phase.

An important question is how to characterize the leptonic
CP-violating effects in neutrino oscillation experiments. For
neutrino oscillations in vacuum, it is evident that the CP
asymmetry, usually defined as $A^{\rm CP}_{\alpha \beta} \equiv
P^{}_{\alpha \beta} - \bar{P}^{}_{\alpha \beta} \propto \sin \delta$
for $\alpha \neq \beta$, is well determined by $\delta$, where
$P^{}_{\alpha \beta} = P(\nu^{}_\alpha \to \nu^{}_\beta)$ and
$\bar{P}^{}_{\alpha \beta} = P(\bar{\nu}^{}_\alpha \to
\bar{\nu}^{}_\beta)$ stand for the neutrino and antineutrino
transition probabilities, respectively. For long-baseline neutrino
oscillation experiments, however, matter effects
\cite{Wolfenstein:1977ue,Mikheev:1986gs} can be significant and
induce fake CP violation, since the Earth matter itself is CP
asymmetric. In this case, the intrinsic CP violation due to $\delta$
in $A^{\rm CP}_{\alpha \beta}$ is obscured by extrinsic CP violation
caused by matter effects. On the other hand, one can determine
$\delta$ by just measuring the probabilities $P^{}_{\alpha \beta}$
as precisely as possible. In this case, if one defines $\Delta
P^{\rm CP}_{\alpha \beta}(\delta) \equiv P^{}_{\alpha \beta}(\delta)
- P^{}_{\alpha \beta}(0)$, the fake CP violation can be removed,
since $\Delta P^{\rm CP}_{\alpha \beta}(\delta)$ vanishes for
$\delta = 0$. Furthermore, it has been suggested
\cite{Kimura:2006jj} that $\Delta P^{\rm m}_{\alpha \beta} \equiv
\max [P^{}_{\alpha \beta}(\delta)] - \min [P^{}_{\alpha
\beta}(\delta)]$ can be utilized to quantify the experimental
sensitivity to $\delta$, where the maximum and minimum are obtained
by freely varying $\delta$ in $[0, 2\pi)$. Both $\Delta P^{\rm
CP}_{\alpha \beta}(\delta)$ and $\Delta P^{\rm m}_{\alpha \beta}$
have been studied in detail by using neutrino oscillograms of the
Earth \cite{Akhmedov:2006hb,Akhmedov:2008qt}.

Since the description of leptonic CP violation should reflect the
difference between neutrinos and antineutrinos, we suggest $\Delta
A^{\rm CP}_{\alpha \beta}(\delta) \equiv A^{\rm CP}_{\alpha
\beta}(\delta) - A^{\rm CP}_{\alpha \beta}(0)$ and $\Delta A^{\rm
m}_{\alpha \beta} \equiv \max[A^{\rm CP}_{\alpha \beta}(\delta)] -
\min[A^{\rm CP}_{\alpha \beta}(\delta)]$ as working observables to
signify the intrinsic CP violation and the experimental sensitivity
to $\delta$. First, we make a general comparison among all five
quantities (i.e., $A^{\rm CP}_{\alpha \beta}$, $\Delta P^{\rm
CP}_{\alpha \beta}$, $\Delta P^{\rm m}_{\alpha \beta}$, $\Delta
A^{\rm CP}_{\alpha \beta}$, and $\Delta A^{\rm m}_{\alpha \beta}$)
and point out their advantages and disadvantages in describing
leptonic CP violation. Then, we perform a detailed numerical study
of them, and describe their main features by using approximate
analytical formulas. Finally, we investigate the experimental setup
that is optimal for the determination of $\delta$.

\section{Measures of leptonic CP violation}

First of all, we briefly review the formulation of three-flavor
neutrino oscillations in matter, which is relevant for long-baseline
experiments. The flavor transition of neutrinos propagating in
matter is governed by the effective Hamiltonian $H^{}_{\rm eff} =
H^{}_{\rm v} + V$, where $H^{}_{\rm v} = U \cdot {\rm diag}(0,
\Delta^{}_{21}, \Delta^{}_{31}) \cdot U^\dagger$ with
$\Delta^{}_{21} \equiv \Delta m^2_{21}/2E$ and $\Delta^{}_{31}
\equiv \Delta m^2_{31}/2E$ being the low and high oscillation
frequencies and $E$ is the neutrino energy. The matter potential is
$V \equiv \sqrt{2} G^{}_{\rm F} n^{}_e {\rm diag}(1, 0, 0)$, where
$n^{}_e$ is the electron number density. In the standard
parametrization, the leptonic mixing matrix is $U = O^{}_{23}
\Gamma_\delta O^{}_{13} \Gamma^\dagger_\delta O^{}_{12}$, where
$O^{}_{ij}$ denotes the rotation in the $i$-$j$ plane with an angle
$\theta^{}_{ij}$ (for $ij = 12, 13, 23$) and $\Gamma^{}_\delta =
{\rm diag}(1, 1, e^{{\rm i}\delta})$ with $\delta$ being the
leptonic Dirac CP-violating phase. Since $V$ is invariant under any
rotations in the $2$-$3$ plane, one can transform to flavor basis
$(\nu^{}_e, \tilde{\nu}^{}_\mu, \tilde{\nu}^{}_\tau)^T =
U^\dagger_{23} (\nu^{}_e, \nu^{}_\mu, \nu^{}_\tau)^T$ with
$U^{}_{23} \equiv O^{}_{23} \Gamma^{}_\delta$ such that
$H^\prime_{\rm eff} = U^\dagger_{23} H^{}_{\rm eff} U^{}_{23}$ is
independent of $\theta^{}_{23}$ and $\delta$ in this basis. The
amplitude of neutrino flavor transition is $A(\nu^{}_\alpha \to
\nu^{}_\beta) = S^{}_{\beta \alpha}$, where the evolution matrix
$S(x)$ satisfies the Schr\"{o}dinger-like equation ${\rm i} [{\rm
d}S(x)/{\rm d}x] = H^{}_{\rm eff}(x) S(x)$ with the initial
condition $S(0) = 1$. For constant matter density, we have $S(x) =
\exp(-{\rm i} H^{}_{\rm eff} x)$ with $x$ being the distance that
neutrinos propagate. If the evolution matrix corresponding to
$H^\prime_{\rm eff}$ is denoted as $S^\prime$, we have $S =
U^{}_{23} S^\prime U^\dagger_{23}$. The oscillation probabilities of
neutrinos are given by $P^{}_{\alpha \beta} = |A(\nu^{}_\alpha \to
\nu^{}_\beta)|^2 = |S^{}_{\beta \alpha}|^2$, while those of
antineutrinos $\bar{P}^{}_{\alpha \beta}$ can be derived from the
same effective Hamiltonian but with the replacements $\delta \to
-\delta$ and $V \to -V$.

As shown for example in Ref.~\cite{Akhmedov:2004ny}, two out of nine
probabilities $P^{}_{\alpha \beta}$ are independent. Now, we choose
them as $P^{}_{\mu e}$ and $P^{}_{\mu \mu}$. Our choice of
$P^{}_{\mu e}$ and $P^{}_{\mu \mu}$ has two advantages: (1) Most of
the neutrino detectors are designed for detection of electrons and
muons and their antiparticles; (2) Both appearance and disappearance
channels are included, which have very different sensitivities to
the neutrino mass hierarchy and $\delta$.

We further show the dependence of $P^{}_{\mu e}$ and $P^{}_{\mu
\mu}$ on $\delta$. Since $S = U^{}_{23} S^\prime U^\dagger_{23}$,
where $S^\prime$ is independent of $\theta^{}_{23}$ and $\delta$, it
is straightforward to prove that $P^{}_{\mu e}$ and $P^{}_{\mu \mu}$
can be written as \cite{Kimura:2002hb,Kimura:2002wd}
\begin{eqnarray}
\label{eq:exact} P^{}_{\mu e} &=& a \cos \delta + b \sin \delta
+ c \; , \nonumber \\
P^{}_{\mu \mu} &=& f \cos \delta + g \cos 2\delta + h \; ,
\end{eqnarray}
where the relevant coefficients $\{a, b, c\}$ and $\{f, g, h\}$ are
independent of $\delta$ and their exact expressions can be found in
Ref.~\cite{Kimura:2002wd}. Using the two independent probabilities
in Eq.~(\ref{eq:exact}), one can readily find the exact expressions
for all the other probabilities.

For neutrino oscillations in vacuum, the measure of leptonic CP
violation can be taken to be the Jarlskog invariant ${\cal J} \equiv
s^{}_{12}c^{}_{12} s^{}_{23} c^{}_{23} s^{}_{13} c^2_{13} \sin
\delta$, where $s^{}_{ij} \equiv \sin \theta^{}_{ij}$ and $c^{}_{ij}
\equiv \cos \theta^{}_{ij}$. In fact, the difference between
neutrino and antineutrino probabilities is $P^{}_{\alpha \beta} -
\bar{P}^{}_{\alpha \beta} \propto {\cal J}$ for $\alpha \neq \beta$.
For neutrino oscillations in matter, however, the situation is
complicated by the CP-asymmetric medium, since only particles rather
than antiparticles are present in Earth matter. In this case, the
discrepancy between $P^{}_{\alpha \beta}$ and $\bar{P}^{}_{\alpha
\beta}$ receives contributions both from $\delta$ and matter
effects. Therefore, it is natural to ask which measure is the best
to extract information on $\delta$ from observations, and to find
the optimal experimental setup to measure $\delta$ in future
neutrino experiments. In the literature, there already exist three
distinct measures:

(i) $A^{\rm CP}_{\alpha \beta}(\delta) \equiv P^{}_{\alpha
\beta}(\delta) - \bar{P}^{}_{\alpha \beta}(\delta)$ denotes the
differences between neutrino probabilities $P^{}_{\alpha
\beta}(\delta)$ and antineutrino probabilities $\bar{P}^{}_{\alpha
\beta}(\delta)$, where the dependence on $\delta$ is explicitly
displayed. In long-baseline experiments, where matter effects play
an important role, $A^{\rm CP}_{\alpha \beta}(\delta)$ is no longer
the best measure of intrinsic CP violation, since $A^{\rm
CP}_{\alpha \beta}(\delta) \neq 0$ even for $\delta = 0$ due to
matter effects. With help of the exact formula for $P^{}_{\mu e}$ in
Eq.~(\ref{eq:exact}) as well as the counterpart $\bar{P}^{}_{\mu e}$
with coefficients $\{\bar{a}, \bar{b}, \bar{c}\}$ in the
antineutrino channel, we obtain
\begin{eqnarray}
\label{eq:Acp} A^{\rm CP}_{\mu e}(\delta) &=& \Delta a \cos \delta +
\Delta b \sin \delta + \Delta c \; ,
\end{eqnarray}
where $\Delta a \equiv a - \bar{a}$; likewise for $\Delta b$ and
$\Delta c$. Obviously, $A^{\rm CP}_{\mu e}(\delta)$ follows the same
dependence on $\delta$ as $P^{}_{\mu e}(\delta)$ and
$\bar{P}^{}_{\mu e}(\delta)$. Note that we focus on the appearance
channel $\nu^{}_\mu (\bar{\nu}^{}_\mu) \to \nu^{}_e
(\bar{\nu}^{}_e)$, but the disappearance channel $\nu^{}_\mu
(\bar{\nu}^{}_\mu) \to \nu^{}_\mu (\bar{\nu}^{}_\mu)$ can be
discussed in a similar way.

(ii) $\Delta P^{\rm CP}_{\alpha \beta}(\delta) \equiv P^{}_{\alpha
\beta}(\delta) - P^{}_{\alpha \beta}(0)$ denotes the differences
between the probabilities $P^{}_{\alpha \beta}$ for an arbitrary
$\delta$ and those for $\delta = 0$. This measure is intended to
remove the fake CP violation induced by matter effects, which has
the advantage that only the neutrino channel is involved. Using
Eq.~(\ref{eq:exact}), we find
\begin{equation}
\label{eq:DeltaPcp} \Delta P^{\rm CP}_{\mu e}(\delta) = 2 \sqrt{a^2
+ b^2}\sin\frac{\delta}{2} \sin \left(\omega -
\frac{\delta}{2}\right)
\end{equation}
with $\tan \omega = b/a$. It is now evident that $\Delta P^{\rm
CP}_{\mu e}(\delta)$ is proportional to $\sin(\delta/2)$, and
vanishes for $\delta = 0$.

(iii) $\Delta P^{\rm m}_{\alpha \beta} \equiv \max[P^{}_{\alpha
\beta}(\delta)] - \min[P^{}_{\alpha \beta}(\delta)]$ denotes the
variation of the probabilities $P^{}_{\alpha \beta}(\delta)$ for
$\delta$ varying in $[0, 2\pi)$. Such a measure is actually
independent from the true value of $\delta$, which is yet unknown,
so it should be useful to find an optimal experimental setup that is
most sensitive to $\delta$. In fact, we have
\begin{eqnarray}
\label{eq:DeltaPm} \Delta P^{\rm m}_{\mu e} &=& 2\sqrt{a^2 + b^2}
\;.
\end{eqnarray}
Hence, one can observe that $\Delta P^{\rm m}_{\mu e}$ essentially
determines the amplitude of $\Delta P^{\rm CP}_{\mu e}(\delta)$. The
basic strategy to probe $\delta$ may first be to optimize the
experimental setup with help of $\Delta P^{\rm m}_{\mu e}$, and then
to extract $\delta$ from the observation of $\Delta P^{\rm CP}_{\mu
e}(\delta)$. In this sense, $\Delta P^{\rm m}_{\mu e}$ and $\Delta
P^{\rm CP}_{\mu e}(\delta)$ can be grouped as a pair of working
observables to probe $\delta$. Both of them are based on $P^{}_{\mu
e}$. Similarly, one can consider $\Delta \bar{P}^{\rm m}_{\mu e}$
and $\Delta \bar{P}^{\rm CP}_{\mu e}(\delta)$ in the antineutrino
channel.

We argue that the proper measures of leptonic CP violation should
manifest the difference between neutrinos and antineutrinos. In
principle, most of the proposed long-baseline experiments are
equally operative in the neutrino and antineutrino channels.
Therefore, based on Eq.~(\ref{eq:Acp}), we suggest a new pair of
working observables:

(iv) $\Delta A^{\rm CP}_{\alpha \beta}(\delta) \equiv A^{\rm
CP}_{\alpha \beta}(\delta) - A^{\rm CP}_{\alpha \beta}(0)$ signifies
the intrinsic CP violation, compared to $A^{\rm CP}_{\alpha
\beta}(\delta)$. A similar quantity, but with a different
normalization, was previously considered
\cite{Donini:1999jc,Altarelli:2008yr}. Using the exact expression
for $A^{\rm CP}_{\mu e}(\delta)$ in Eq.~(\ref{eq:Acp}), we arrive at
\begin{equation}
\label{eq:DeltaAcp} \Delta A^{\rm CP}_{\mu e}(\delta) = 2
\sqrt{(\Delta a)^2 + (\Delta b)^2} \sin\frac{\delta}{2} \sin
\left(\omega^\prime - \frac{\delta}{2}\right) ~~
\end{equation}
with $\tan \omega^\prime = \Delta b/\Delta a$. Note that this result
takes the same form as that of $\Delta P^{\rm CP}_{\mu e}(\delta)$
in Eq.~(\ref{eq:DeltaPcp}), except for the relevant coefficients.

(v) $\Delta A^{\rm m}_{\alpha \beta} \equiv \max[A^{\rm CP}_{\alpha
\beta}(\delta)] - \min[A^{\rm CP}_{\alpha \beta}(\delta)]$ denotes
the variation of $A^{\rm CP}_{\alpha \beta}(\delta)$ for $\delta$
varying in $[0, 2\pi)$. The extrinsic CP-violating effects cancel in
$\Delta A^{\rm m}_{\alpha \beta}$. It is straightforward to show
that
\begin{equation}
\label{eq:DeltaAm} \Delta A^{\rm m}_{\mu e} = 2 \sqrt{(\Delta a)^2 +
(\Delta b)^2} \;.
\end{equation}
The size of $\Delta A^{\rm m}_{\mu e}$ determines the
magnitude of $\Delta A^{\rm CP}_{\mu e}(\delta)$ through the
$\delta$-independent coefficient.

We expect that $\Delta A^{\rm CP}_{\alpha \beta}$ and $\Delta A^{\rm
m}_{\alpha \beta}$ can be implemented to extract $\delta$, and to
optimize the experimental setup, similar to $\Delta P^{\rm
CP}_{\alpha \beta}$ and $\Delta P^{\rm m}_{\alpha \beta}$.
Nevertheless, the former ones contain the difference between
neutrino and antineutrino probabilities, so these two sets of
measures are not equivalent. In the following, we will present
approximate and analytical results for all the above measures of
leptonic CP violation, and the numerical results as well.
Furthermore, the optimal experimental setup for probing $\delta$ is
considered and compared with the ongoing and upcoming neutrino
oscillation experiments.

\section{Analytical \& Numerical Results}

We define $\alpha \equiv \Delta^{}_{21}/\Delta^{}_{31}$, $\Delta
\equiv \Delta^{}_{31} L/2$ with $L$ being the distance between the
source and detector, and $A \equiv V/\Delta^{}_{31}$, where $\alpha$
denotes the ratio of the low and high oscillation frequencies,
$\Delta$ is the phase corresponding to the high oscillation
frequency, and $A$ stands for the strength of matter effects. The
probabilities in matter of constant density have been calculated in
Ref.~\cite{Akhmedov:2004ny} to the second order in both $\alpha$ and
$s^{}_{13}$. According to the current neutrino oscillation data, we
have $\alpha \approx \sqrt{2} s^2_{13} \approx 0.03$
\cite{Fogli:2012ua,Tortola:2012te,GonzalezGarcia:2012sz}. Therefore,
one can expand the probabilities in terms of $\alpha$ and
$s^{}_{13}$, and neglect all higher-order terms of ${\cal
O}(\alpha^2)$. Using the approximate formula for $P^{}_{\mu e}$, we
can identify the corresponding coefficients
\begin{eqnarray}
\label{eq:abc} a &\approx& +8\alpha {\cal J}^{}_{\rm r} \frac{\sin A
\Delta}{A} \frac{\sin (A-1) \Delta}{A-1} \cos \Delta \;,
\nonumber \\
b &\approx& -8\alpha {\cal J}^{}_{\rm r} \frac{\sin A \Delta}{A}
\frac{\sin (A-1) \Delta}{A-1} \sin \Delta \;,
\nonumber \\
c &\approx& 4s^2_{13} s^2_{23} \frac{\sin^2 (A-1) \Delta}{(A-1)^2}
\;
\end{eqnarray}
with ${\cal J}^{}_{\rm r} \equiv {\cal J}/\sin \delta \approx
s^{}_{13} s^{}_{12} c^{}_{12} s^{}_{23} c^{}_{23}$ being the reduced
Jarlskog invariant. The coefficients $\{\bar{a}, \bar{b}, \bar{c}\}$
for $\bar{P}^{}_{\mu e}$ can be obtained by replacing $A$ with $-A$
and $\delta$ with $-\delta$ in $P^{}_{\mu e}$. Hence, we obtain
\begin{eqnarray}
\label{eq:Deltaabc} \Delta a &\approx& +8\alpha {\cal J}^{}_{\rm r}
\Theta^{}_- \frac{\sin A \Delta}{A}
\cos \Delta \; , \nonumber \\
\Delta b &\approx& -8\alpha {\cal J}^{}_{\rm r} \Theta^{}_+
\frac{\sin A \Delta}{A}
\sin \Delta \; , \nonumber \\
\Delta c &\approx& 4s^2_{13} s^2_{23} \Theta^{}_+ \Theta^{}_-
\end{eqnarray}
with $\Theta^{}_\pm \equiv \sin [(A-1)\Delta]/(A-1) \pm \sin
[(A+1)\Delta]/(A+1)$. Note that Eqs.~(\ref{eq:abc}) and
(\ref{eq:Deltaabc}) are valid as long as $\alpha \Delta \ll 1$,
i.e., when the distance $L$ and energy $E$ are far away from the
region where the low-frequency oscillation becomes dominant. This
condition is satisfied in all the ongoing and upcoming long-baseline
experiments, however, it is violated for atmospheric neutrino
experiments. When low-frequency oscillations come into play, one can
expand the probabilities in terms of $s^{}_{13}$, which are exact
with respect to $\alpha$, as is performed in
Ref.~\cite{Akhmedov:2004ny}.

Now, we apply the approximate formulas to the measures of leptonic
CP violation in Eqs.~(\ref{eq:Acp})-(\ref{eq:DeltaAm}) and explore
their main features.

\subsection{CP Asymmetry $A^{\rm CP}_{\mu e}(\delta)$}

%%%%%%%%%%%%%%%%%%%%%%%%%%%    Fig.1  %%%%%%%%%%%%%%%%%%%%%%%%%%%%%
\begin{figure*}
\includegraphics[width=.35\textwidth]{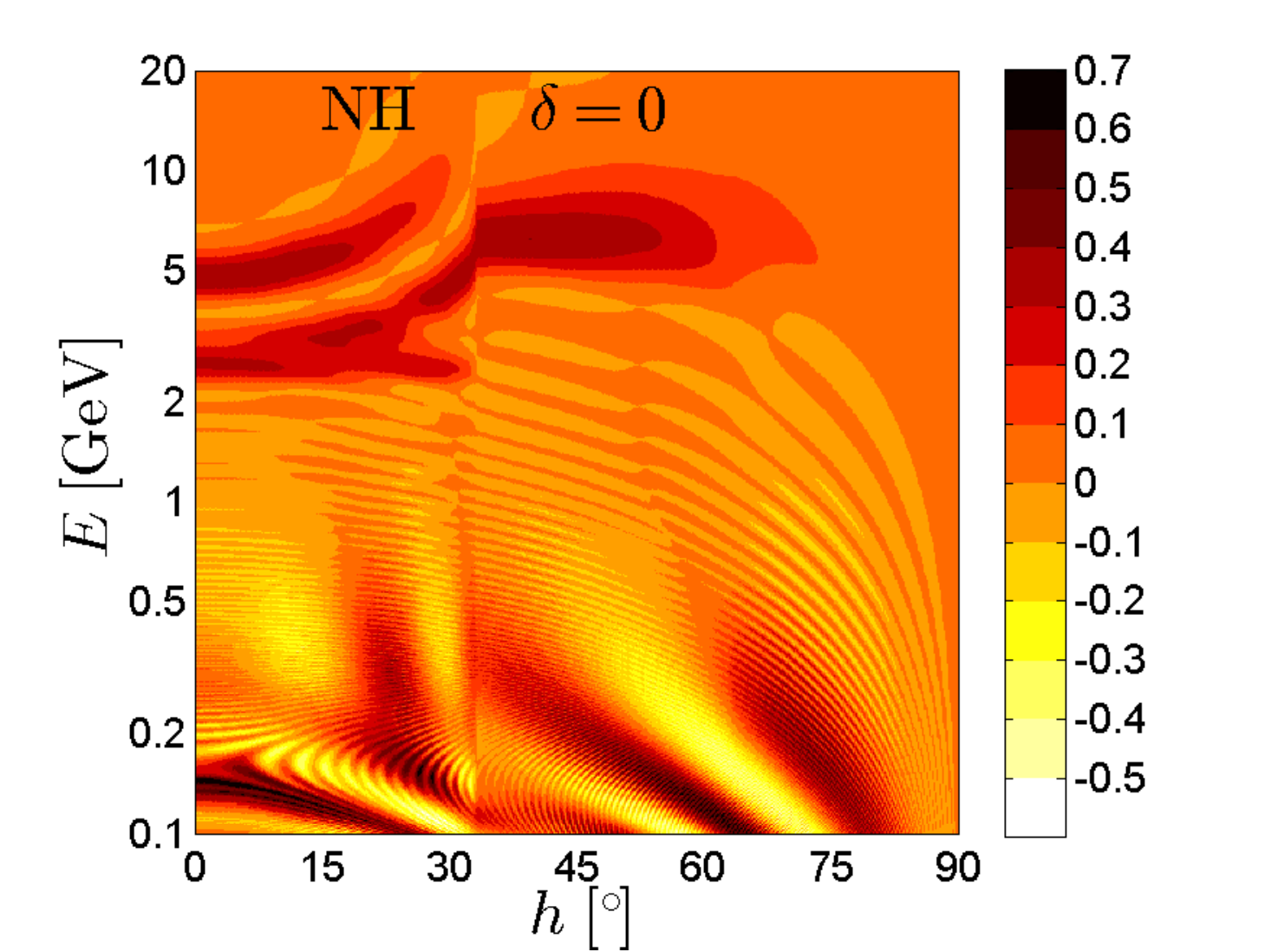}
\includegraphics[width=.35\textwidth]{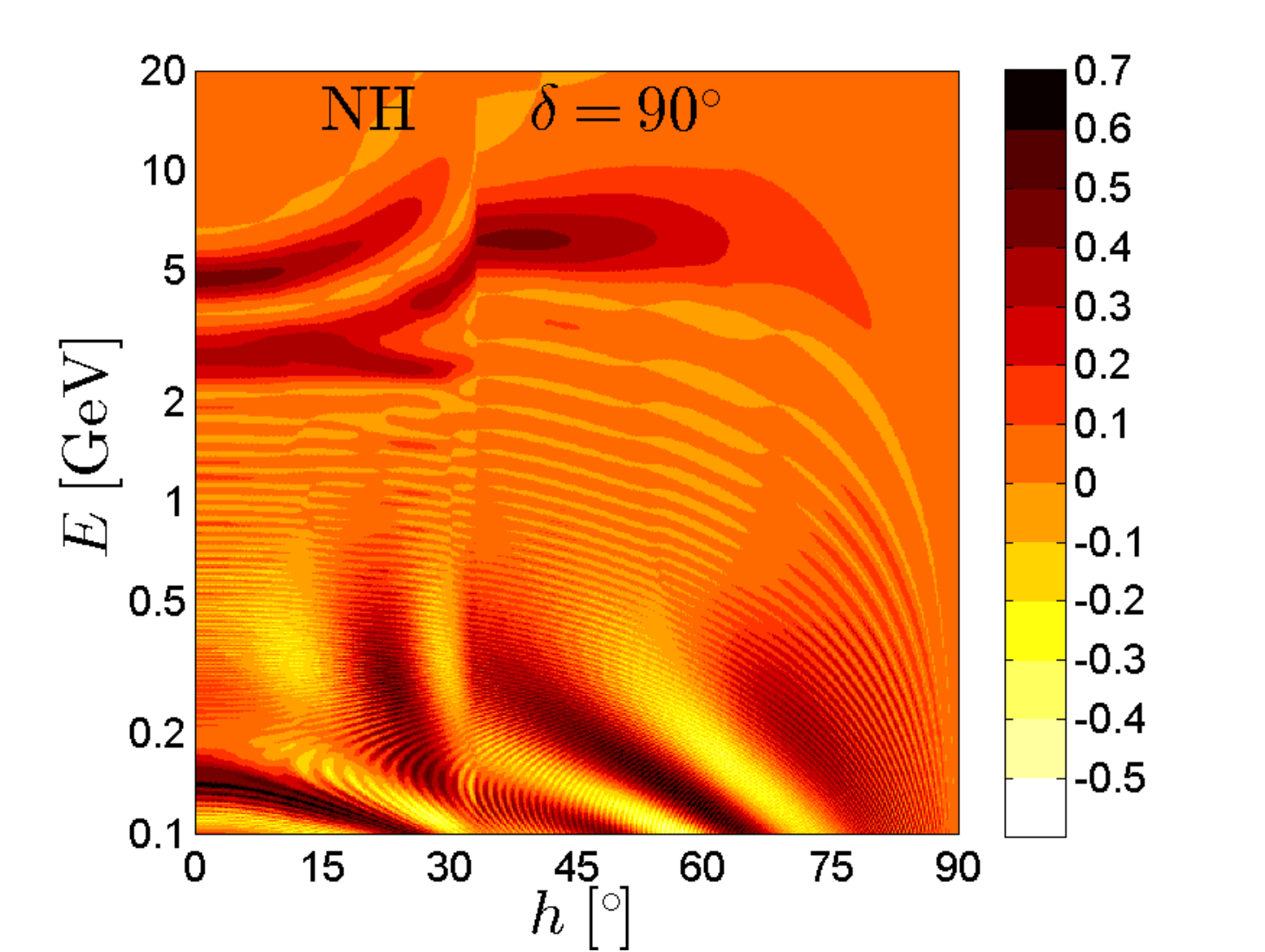}
\includegraphics[width=.35\textwidth]{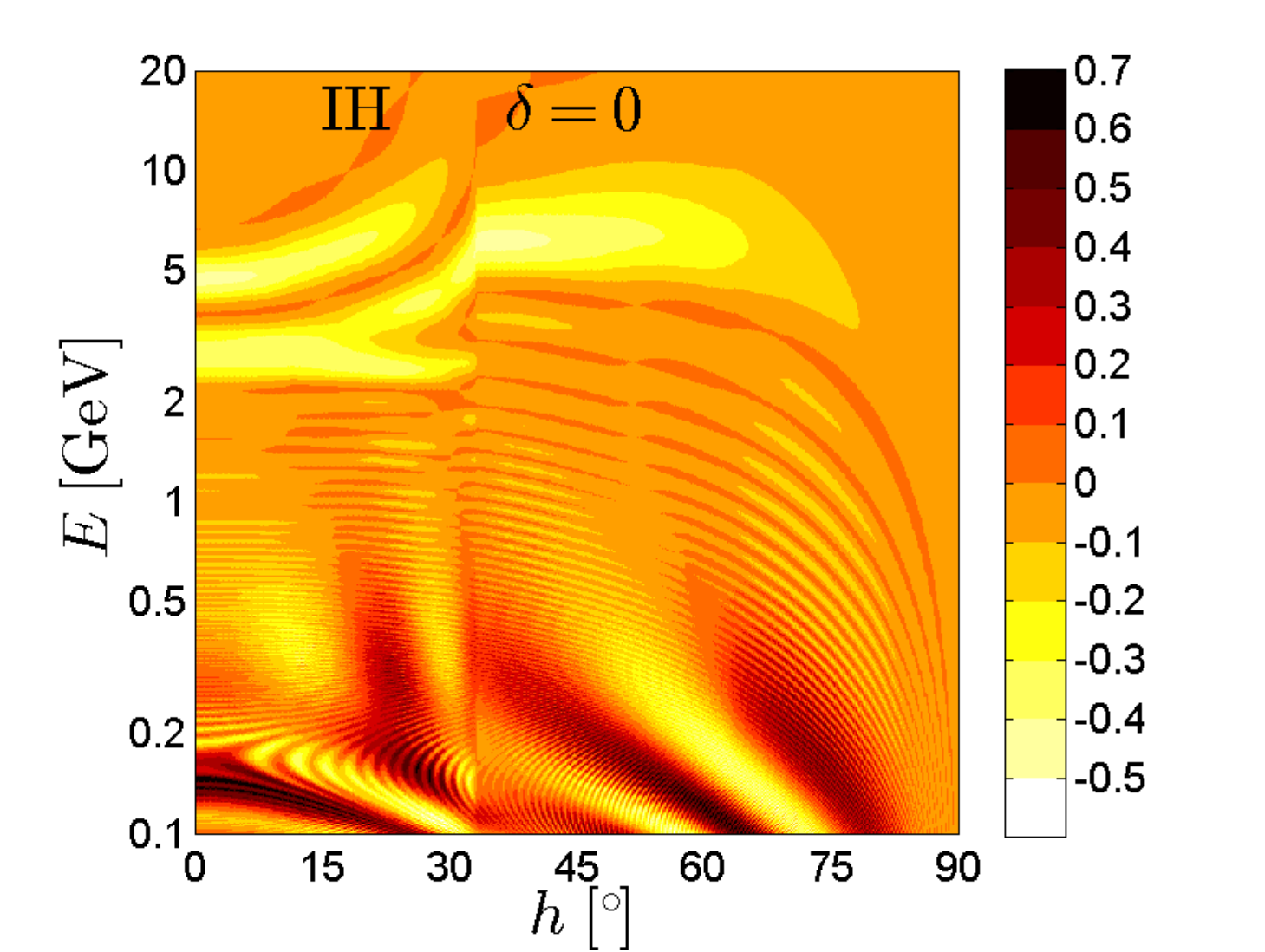}
\includegraphics[width=.35\textwidth]{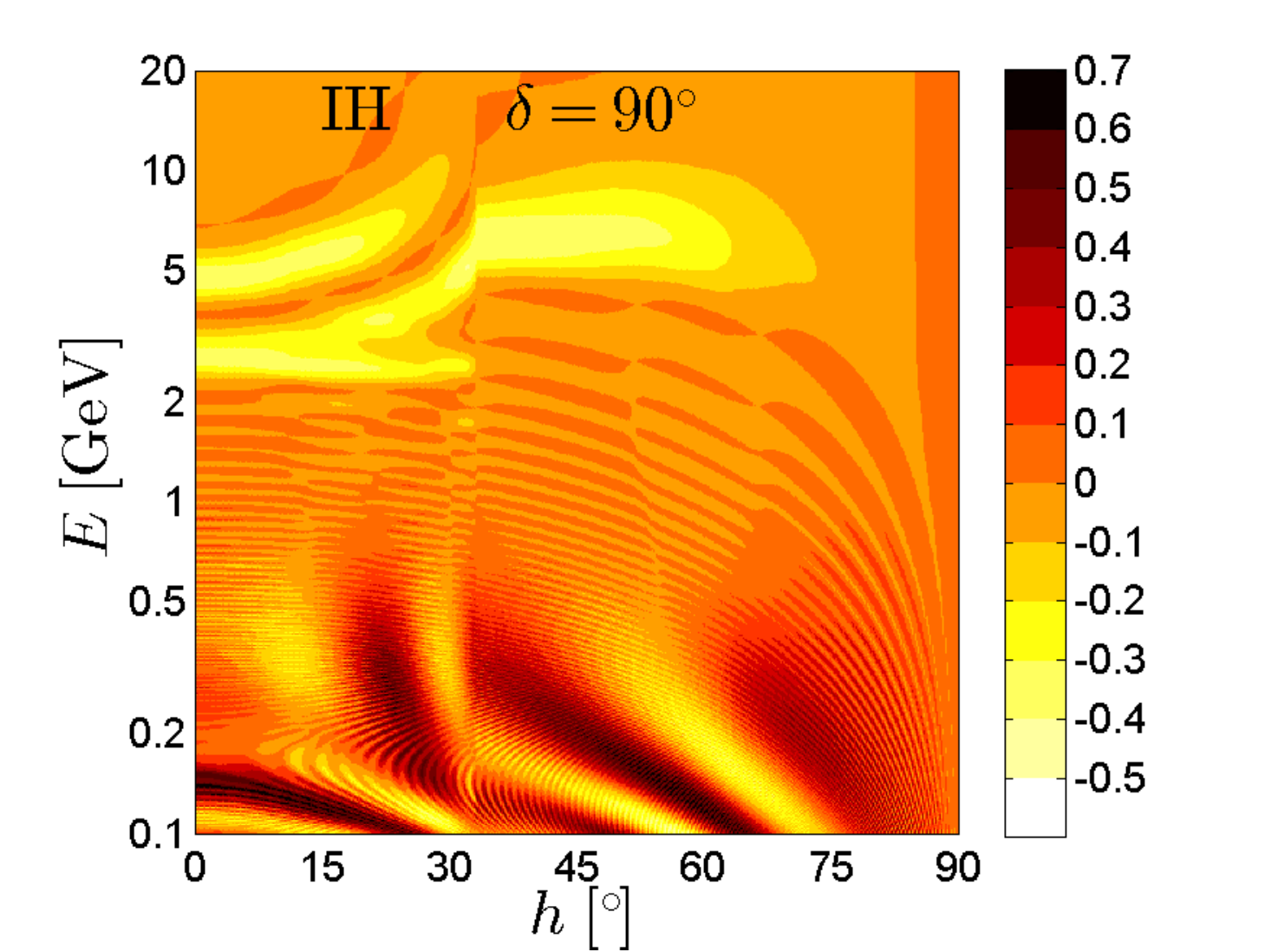}
\caption{Numerical results of $A^{\rm CP}_{\mu e}(\delta)$ for
$\delta = 0$ and $\pi/2$ in the case of normal neutrino mass
hierarchy (upper row) and inverted neutrino mass hierarchy (lower
row), where the best-fit values $\theta^{}_{12} = 34^\circ$,
$\theta^{}_{13} = 9^\circ$, $\theta^{}_{23} = 40^\circ$, $\Delta
m^2_{21} = 7.5\times 10^{-5}~{\rm eV}^2$, and $|\Delta m^2_{31}| =
2.5\times 10^{-3}~{\rm eV}^2$ have been used \cite{Fogli:2012ua}.}
\end{figure*}
%%%%%%%%%%%%%%%%%%%%%%%%%%%%%%%%%%%%%%%%%%%%%%%%%%%%%%%%%%%%%%%%%%
From Eqs.~(\ref{eq:Acp}) and (\ref{eq:Deltaabc}), one can obtain the
approximate result for $A^{\rm CP}_{\mu e}(\delta)$. For neutrino
energies $E > 2~{\rm GeV}$, i.e., in the high-energy approximation,
we can safely ignore $\Delta m^2_{21}$. In this two-flavor limit,
$P^{}_{\mu e}$ is just given by the $\delta$-independent term $c$.
Note that the $\delta$-dependent terms arise from the interference
between the two mass-squared differences, and thus are suppressed by
a factor $\alpha/s^{}_{13}$, indicating that the main structure of
$P^{}_{\mu e}$ in the $L$-$E$ plane is determined by high-frequency
and parametric resonances \cite{Akhmedov:2006hb}. Given the neutrino
mass hierarchy, the resonance existing in the neutrino channel
should be absent in the antineutrino channel, and vice versa. For
lower energies, one has to consider three-flavor oscillations and
analyze the resonance structure due to $\Delta m^2_{21}$. For a
systematic study of neutrino oscillograms of the Earth and the
theoretical interpretation of their resonance structures, see
Refs.~\cite{Akhmedov:2006hb,Akhmedov:2008qt}.

In Fig.~1, we have calculated $A^{\rm CP}_{\mu e}(\delta)$ by using
the exact probabilities in the three-flavor framework and the PREM
model of the Earth matter density \cite{Dziewonski:1981xy}. The
baseline can be calculated via $L = 2R \cos h$ with $R \simeq
6370~{\rm km}$ being the Earth radius and $h$ the nadir angle. The
abrupt transition around $h \approx 33^\circ$ in Fig.~1 is caused by
the change from the core- and mantle-crossing trajectories.

The main structure in Fig.~1 is determined by matter effects. In
particular, the plots for $\delta = 0$ indicate pure fake CP
violation. In high-energy region, we have $A^{\rm CP}_{\mu e}
\approx 4s^2_{13} s^2_{23} \Theta^{}_+ \Theta^{}_-$ to leading
order. In the neutrino channel, the high-frequency resonance exists
for the normal mass hierarchy (NH). In other words, $\sin
(A-1)\Delta$ is resonantly enhanced in the NH case, while $\sin
(A+1)\Delta$ in the inverted mass hierarchy (IH). Therefore, from
the NH case to the IH case, $\Theta^{}_-$ changes sign while
$\Theta^{}_+$ not, which can be used to explain the sign-flipping
difference between the NH and IH cases. In Fig.~1, the numerical
results are presented for the NH and IH cases in the upper and lower
rows, respectively. The sign-flipping difference can be clearly
observed in the high-energy region. In the sub-GeV energy region,
the low-frequency resonances and parametric enhancement comprise the
dominant structure \cite{Akhmedov:2008qt}. The peak and valley
features come from the mismatch between neutrino and antineutrino
probabilities. Hence, the results of $A^{\rm CP}_{\mu e}(\delta)$ in
this region are essentially independent of the different neutrino
mass hierarchies, and tiny differences may arise from the high-order
corrections of $s^{}_{13}$. The difference between the two plots in
each row of Fig.~1 is more evident in the low-energy region, where
the condition for the high-frequency resonance is not fulfilled.

\subsection{Working Observables $\Delta P^{\rm CP}_{\mu e}(\delta)$ and $\Delta P^{\rm m}_{\mu e}$}

%%%%%%%%%%%%%%%%%%%%%%%%%%%    Fig.2  %%%%%%%%%%%%%%%%%%%%%%%%%%%%%
\begin{figure*}
\includegraphics[width=.35\textwidth]{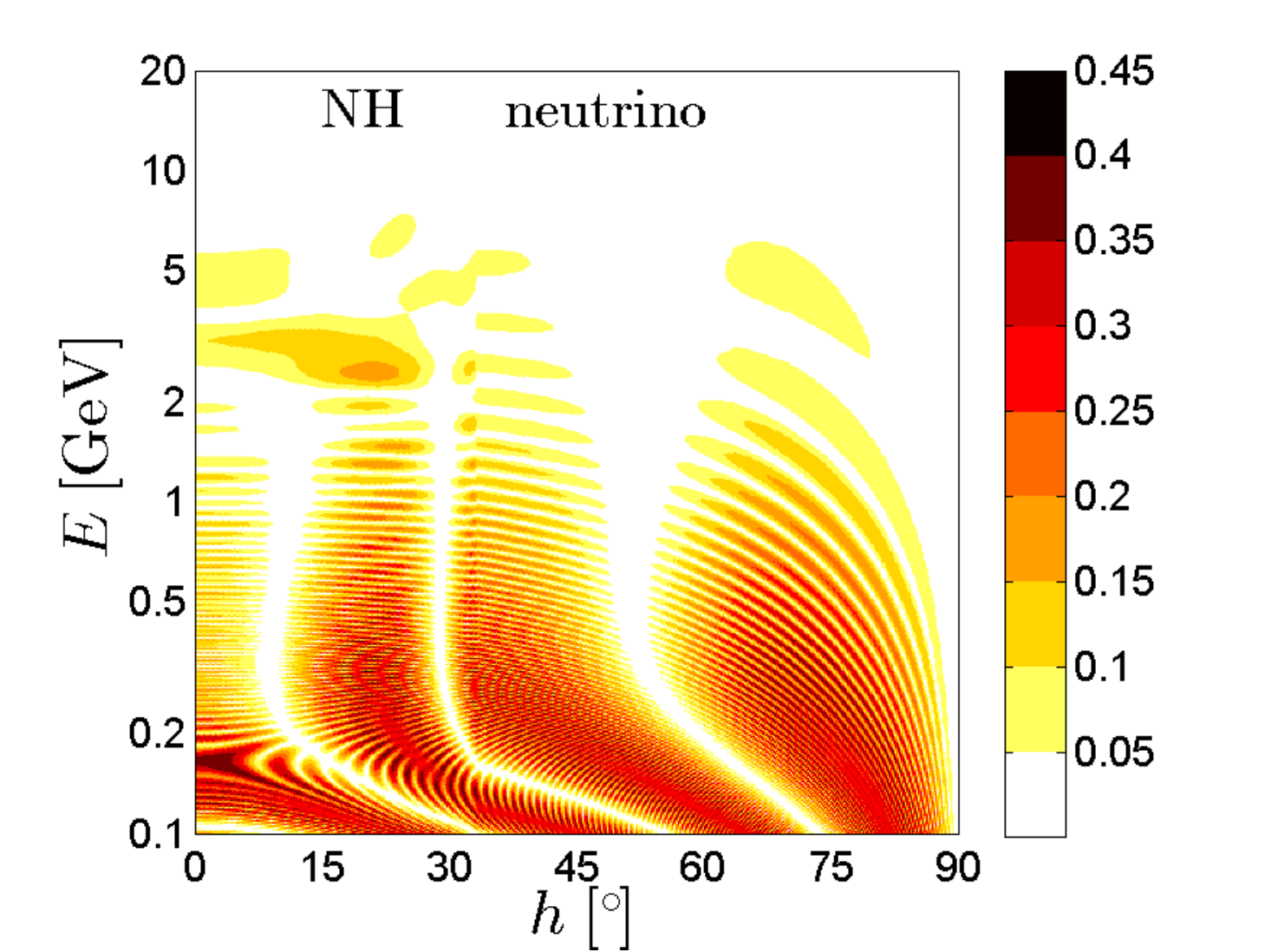}
\includegraphics[width=.35\textwidth]{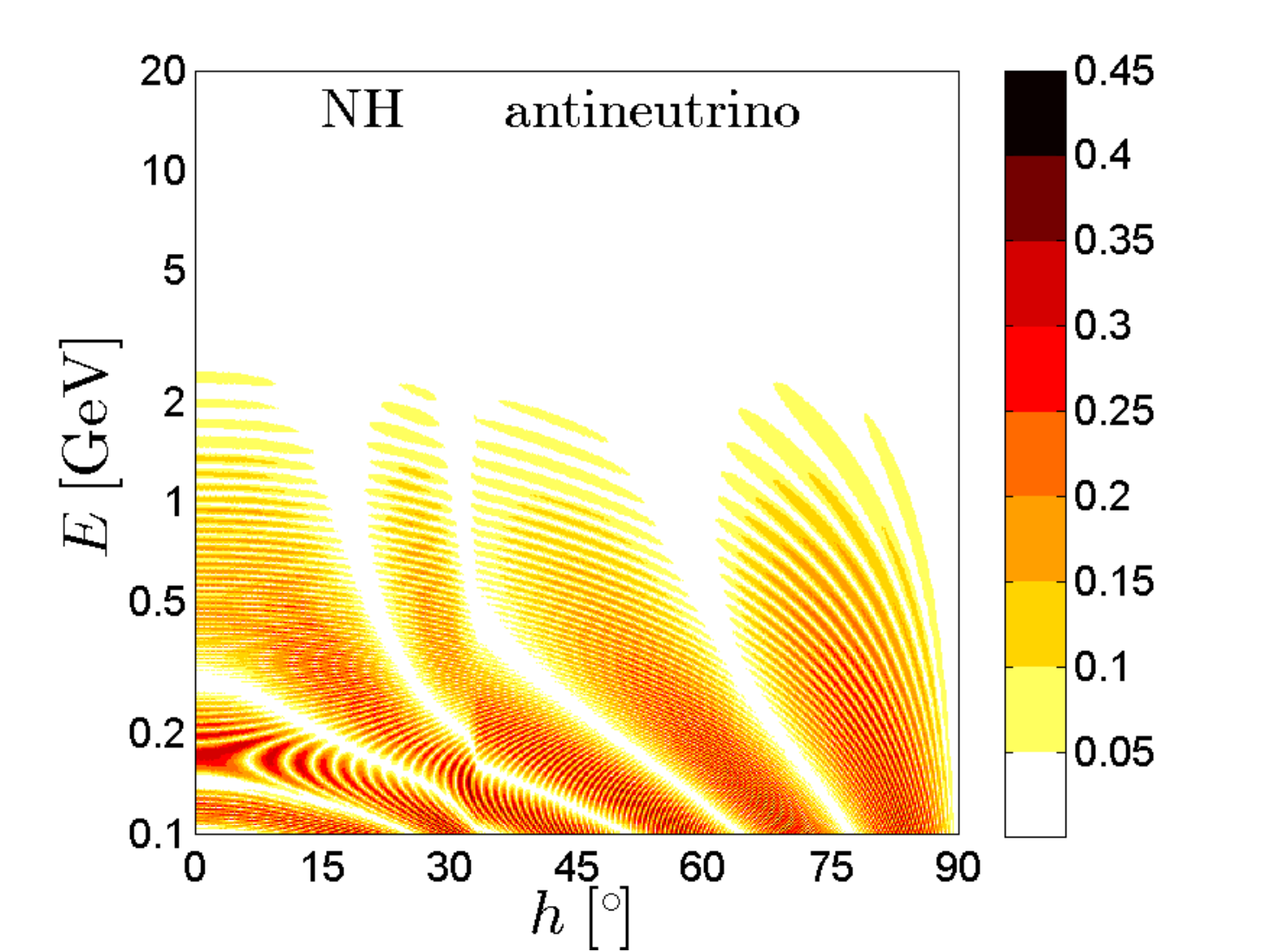}
\includegraphics[width=.35\textwidth]{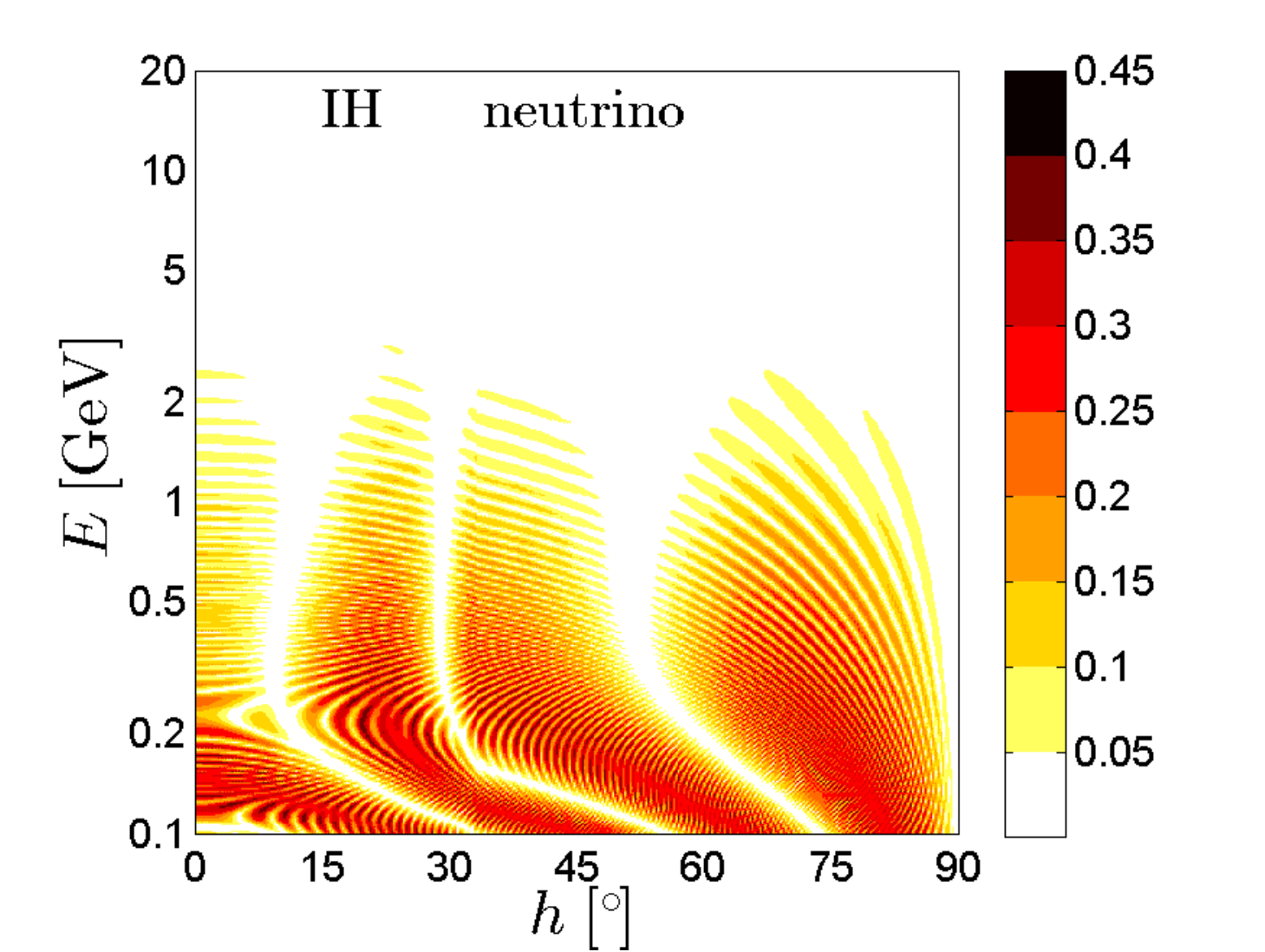}
\includegraphics[width=.35\textwidth]{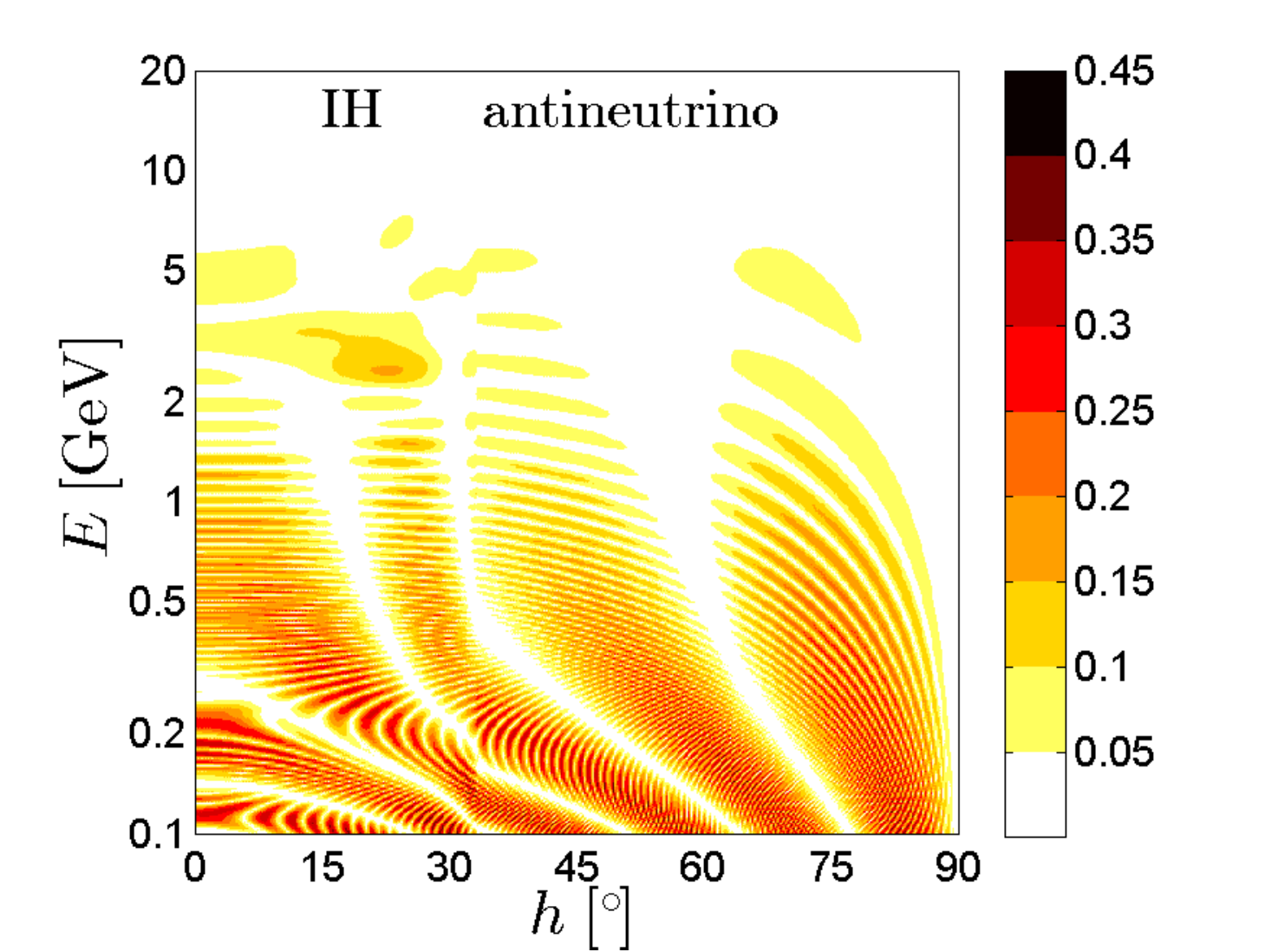}
\caption{Numerical results of $\Delta P^{\rm m}_{\mu e}$ and $\Delta
\bar{P}^{\rm m}_{\mu e}$ in the cases of normal neutrino mass
hierarchy (upper row) and inverted neutrino mass hierarchy (lower
row), where the input neutrino parameters are the same as in
Fig.~1.}
\end{figure*}
%%%%%%%%%%%%%%%%%%%%%%%%%%%%%%%%%%%%%%%%%%%%%%%%%%%%%%%%%%%%%%%%%%

Now, we turn to the working observables $\Delta P^{\rm CP}_{\mu
e}(\delta)$ and $\Delta P^{\rm m}_{\mu e}$ based only on $P^{}_{\mu
e}$. Using Eqs.~(\ref{eq:DeltaPcp}) and (\ref{eq:abc}), we obtain
the approximate formula
\begin{eqnarray}
\Delta P^{\rm CP}_{\mu e} &=& -8 \alpha {\cal J} \frac{\sin
A\Delta}{A} \frac{\sin(A-1)\Delta}{A-1} \nonumber \\
&~& \times \left[\tan\frac{\delta}{2} \cos \Delta + \sin \Delta
\right] \; .
\end{eqnarray}
Note that the Jarlskog invariant ${\cal J}$ appearing here is just a
simple notation and does not mean any difference between neutrino
and antineutrino oscillations.  On the other hand, from
Eqs.~(\ref{eq:DeltaPm}) and (\ref{eq:abc}), we find
\begin{equation}
\Delta P^{\rm m}_{\mu e} =  16\alpha {\cal J}^{}_{\rm r} \frac{\sin
A\Delta}{A} \frac{\sin(A-1)\Delta}{A-1} \; .
\end{equation}

Since both $\Delta P^{\rm CP}_{\mu e}(\delta)$ and $\Delta P^{\rm
m}_{\mu e}$ have already been studied in great detail in
Refs.~\cite{Akhmedov:2006hb,Akhmedov:2008qt}, we will not repeat the
analysis here. However, in Fig.~2, we have performed numerical
calculations of $\Delta P^{\rm m}_{\mu e}$ for neutrinos and $\Delta
\bar{P}^{\rm m}_{\mu e}$ for antineutrinos in both NH and IH cases,
to illustrate the main features.

It is straightforward to understand the similarity between the case
of neutrinos with NH and that of antineutrinos with IH. Note that
$\Delta P^{\rm m}_{\mu e}$ for IH can be obtained by changing $A \to
- A$, $\Delta \to -\Delta$, and $\alpha \to -\alpha$. On the other
hand, the formula of $\Delta \bar{P}^{\rm m}_{\mu e}$ for NH can be
derived by replacing $A \to -A$. One immediately observes that these
two formulas are identical. This feature is only present in the
high-energy region, as shown in Fig.~2, where the similarity between
the case of neutrinos with IH and that of antineutrinos with NH is
also evident.

There is a resonance deep in the core region for NH. But this
resonance ($h \sim 22^\circ$ and $2~{\rm GeV} < E < 3~{\rm GeV}$)
disappears for IH. The oscillatory structures in the low-energy
region are different for neutrinos and antineutrinos, but similar
for NH and IH. Moreover, there are three vertical lines, i.e., the
``solar magic lines" corresponding to $\Delta P^{\rm m}_{\mu e} = 0$
\cite{Akhmedov:2008qt}. Solving $\sin A\Delta = 0$, one obtains $A
\Delta = n \pi$, or $L = 2n\pi/V$, with $n$ being a positive
integer. On the other hand, the equation $\sin (A-1)\Delta = 0$
determines the ``atmospheric magic lines". But these conditions for
magic lines depend on the factorization approximation (i.e., $\alpha
\to 0$ and $s^{}_{13} \to 0$), and the general forms can be found in
Ref.~\cite{Akhmedov:2008qt}. In the low-energy region, the ``solar
magic lines" are no longer energy-independent, and they coincide
with the low-frequency oscillation dips.

\subsection{Working Observables $\Delta A^{\rm CP}_{\mu e}(\delta)$ and $\Delta A^{\rm m}_{\mu e}$}

%%%%%%%%%%%%%%%%%%%%%%%%%%%    Fig.3  %%%%%%%%%%%%%%%%%%%%%%%%%%%%%
\begin{figure*}
\includegraphics[width=.35\textwidth]{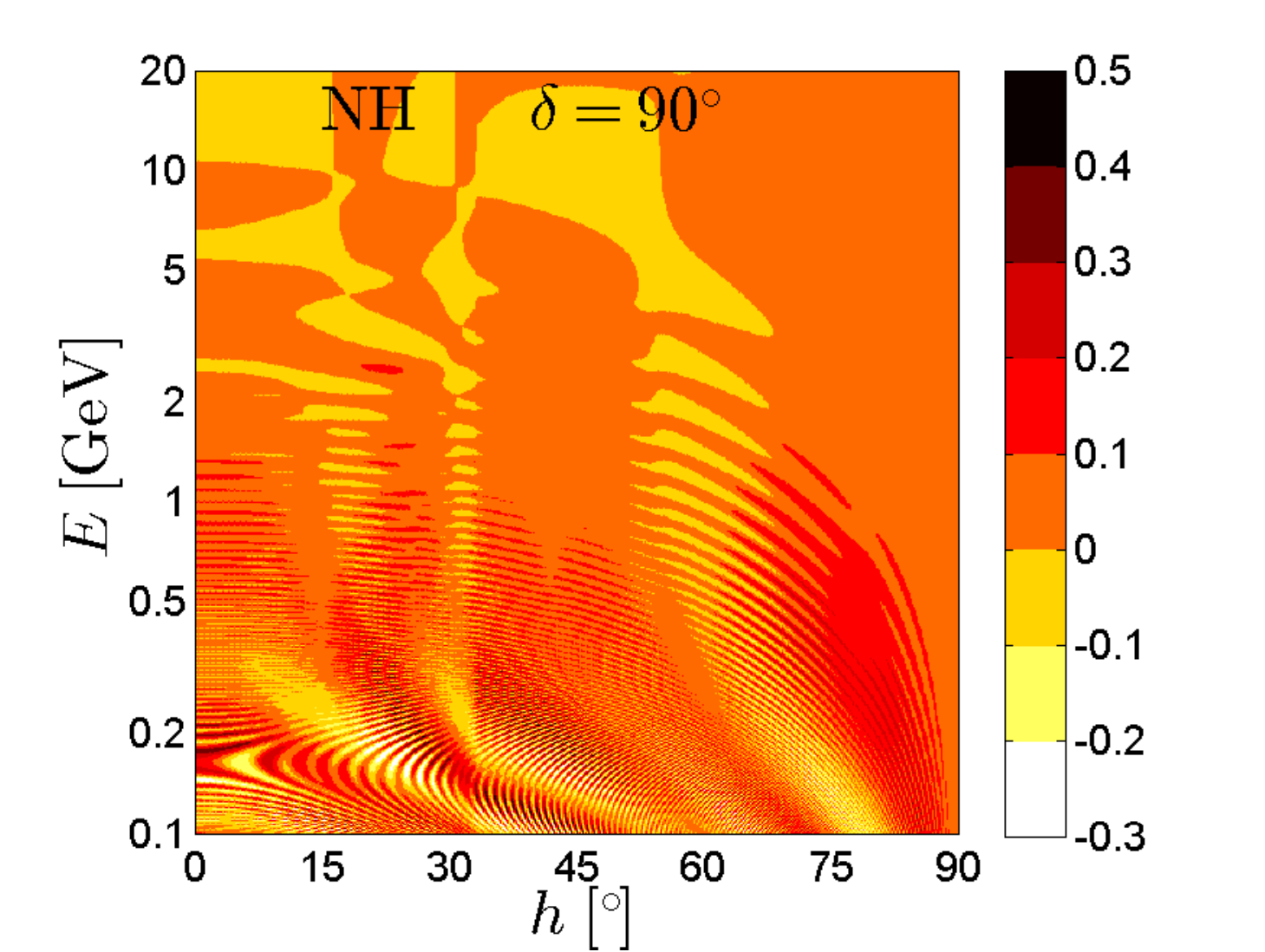}
\includegraphics[width=.35\textwidth]{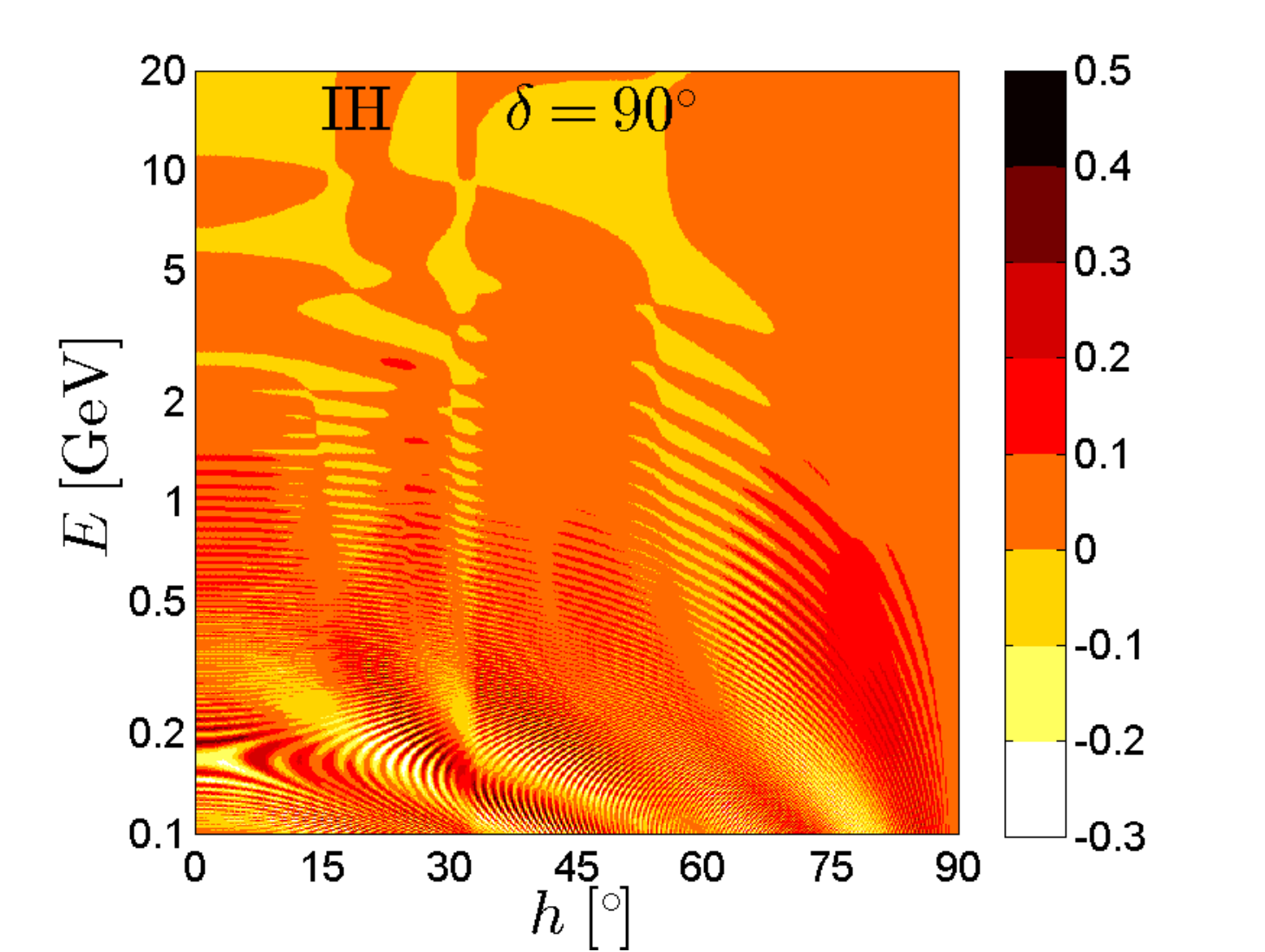}
\caption{Numerical results of $\Delta A^{\rm CP}_{\mu e}(\delta)$
for $\delta = \pi/2$ in the cases of normal neutrino mass hierarchy
(left) and inverted neutrino mass hierarchy (right), where the input
neutrino parameters are the same as in Fig.~1.}
\end{figure*}
%%%%%%%%%%%%%%%%%%%%%%%%%%%%%%%%%%%%%%%%%%%%%%%%%%%%%%%%%%%%%%%%%%

%%%%%%%%%%%%%%%%%%%%%%%%%%%    Fig.4  %%%%%%%%%%%%%%%%%%%%%%%%%%%%%
\begin{figure*}
\includegraphics[width=.35\textwidth]{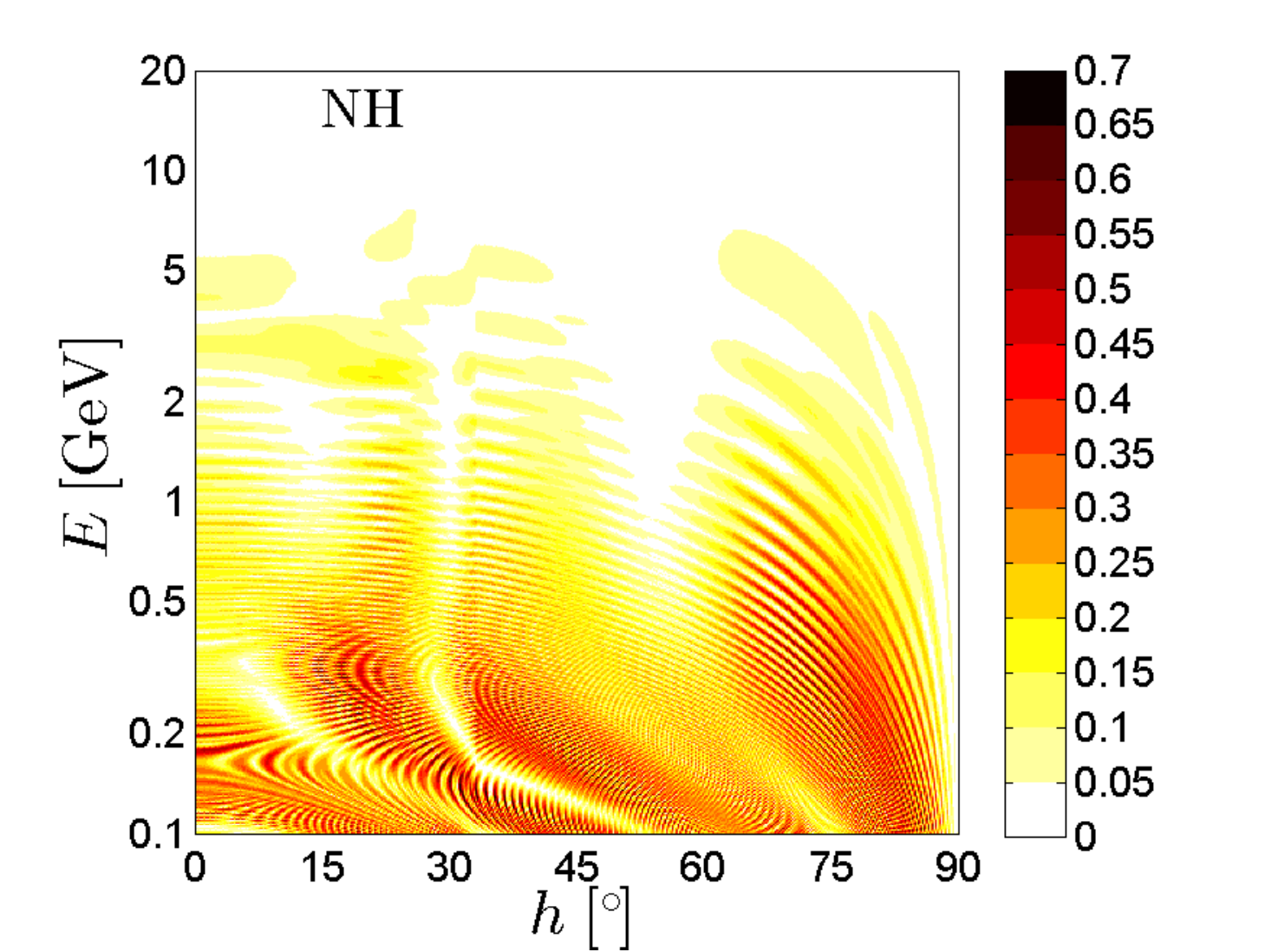}
\includegraphics[width=.35\textwidth]{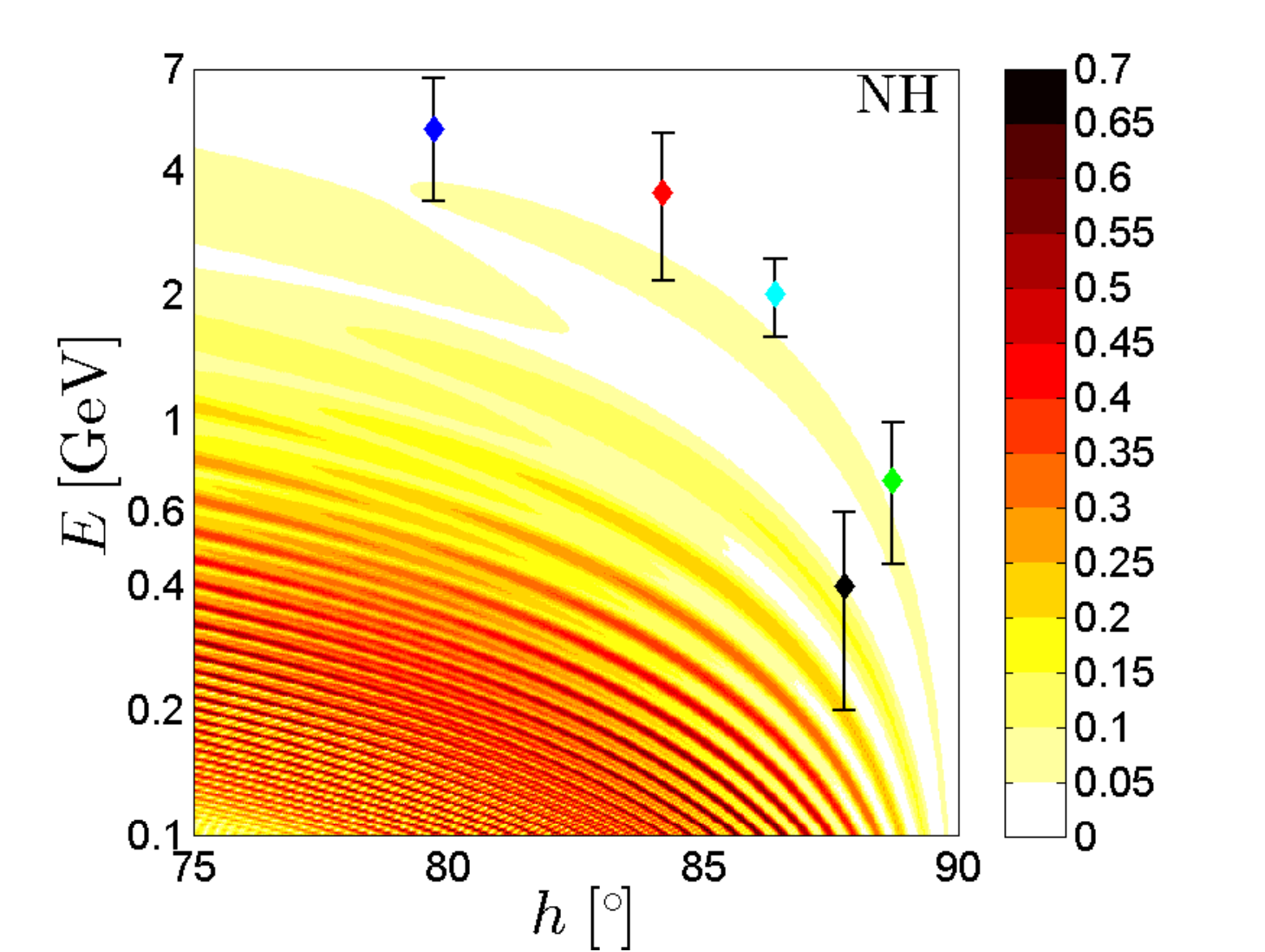}
\includegraphics[width=.35\textwidth]{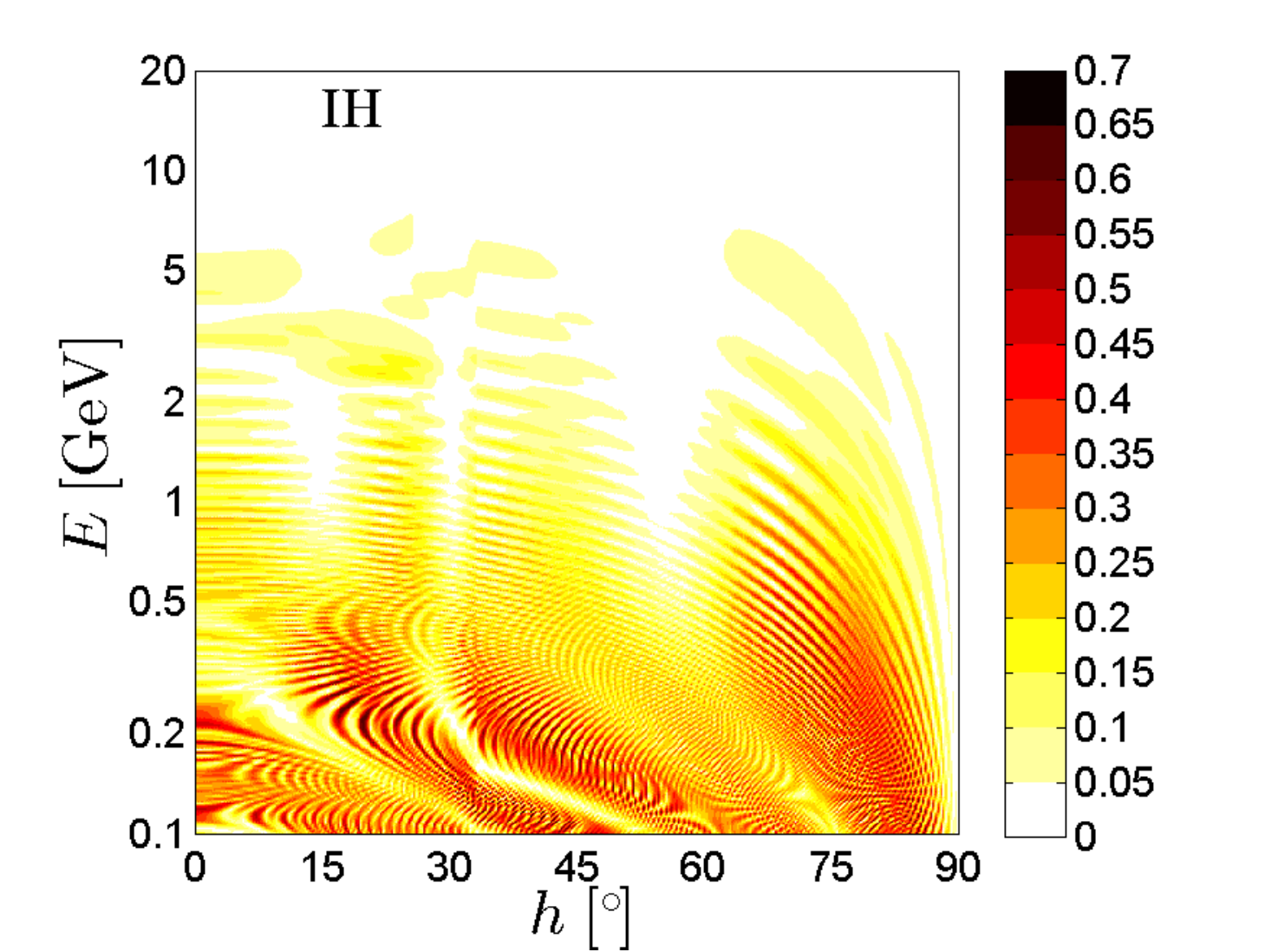}
\includegraphics[width=.35\textwidth]{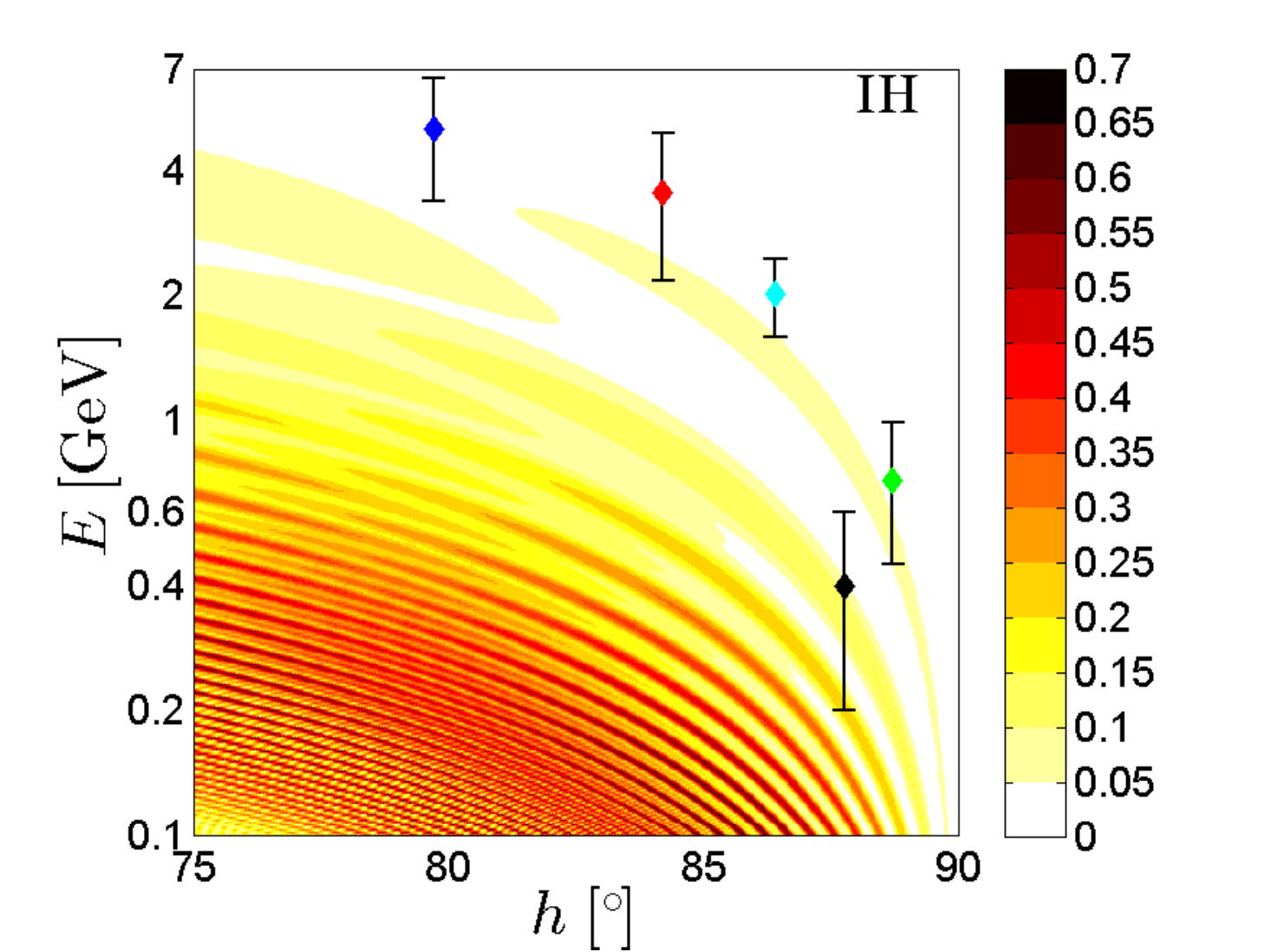}
\caption{Numerical results of $\Delta A^{\rm m}_{\mu e}$ in the case
of normal neutrino mass hierarchy (upper row) and inverted neutrino
mass hierarchy (lower row), where the input neutrino parameters are
the same as in Fig.~1. In the right plots, we have zoomed in the
parameter region that is relevant for the long-baseline neutrino
oscillation experiments T2K (green, at $h = 88.7^\circ$), NO$\nu$A
(cyan, at $h = 86.4^\circ$), LBNE (red, at $h = 84.1^\circ$),
LAGUNA-LBNO (blue, $h = 79.7^\circ$), and ESS (black, at $h =
87.8^\circ$).}
\end{figure*}
%%%%%%%%%%%%%%%%%%%%%%%%%%%%%%%%%%%%%%%%%%%%%%%%%%%%%%%%%%%%%%%%%%
Finally, we come to the new pair of working observables $\Delta
A^{\rm CP}_{\mu e}(\delta)$ and $\Delta A^{\rm m}_{\mu e}$. The
motivation to introduce these observables is two-fold. First, any
quantity measuring leptonic CP violation in neutrino oscillations
should reflect the intrinsic difference between neutrino and
antineutrino probabilities. Second, it should be helpful for an
optimal experimental setup to practically measure $\delta$.
Therefore, we derive the working observables from $A^{\rm CP}_{\mu
e}$. Combining Eq.~(\ref{eq:DeltaAcp}) with Eq.~(\ref{eq:Deltaabc}),
we obtain
\begin{equation}
\Delta A^{\rm CP}_{\mu e} = -8\alpha {\cal J} \frac{\sin A\Delta}{A}
\left[\Theta^{}_-  \tan \frac{\delta}{2} \cos \Delta + \Theta^{}_+
\sin \Delta\right] \; ,
\end{equation}
which reduces to the vacuum result $A^{\rm CP}_{\mu e}$ when $A \to
0$. As $\Delta A^{\rm CP}_{\mu e}$ is proportional to ${\cal J}$, it
vanishes for $\delta = 0$ and the CP-violating effects induced by
matter effects have been removed. In Fig.~3, we show the numerical
results of $\Delta A^{\rm CP}_{\mu e}$ for $\delta = \pi/2$ in both
NH and IH. By definition, they are just the differences between the
two plots in each row of Fig.~1. Apart from a resonance region in
the deep core, sizable values of $\Delta A^{\rm CP}_{\mu e}$ are
lying in the energy region below $1~{\rm GeV}$. Note that the
results for NH and those for IH in the high-energy region are nearly
indistinguishable, which can be understood by noting that Eq.~(11)
is invariant under the transformations $\alpha \to -\alpha$, $\Delta
\to -\Delta$, and $A \to -A$.

To examine the sensitivity to $\delta$, we consider the range of
$A^{\rm CP}_{\mu e}(\delta)$ by varying $\delta$ in $[0,2\pi)$. From
Eqs.~(\ref{eq:DeltaAm}) and (\ref{eq:Deltaabc}), we find
\begin{equation}
\Delta A^{\rm m}_{\mu e} = 16\alpha {\cal J}^{}_{\rm r}
\sqrt{(\Theta^{}_- \cos\Delta)^2+(\Theta^{}_+
\sin\Delta)^2}\frac{\sin A\Delta}{A} \; .
\end{equation}
Similar to $\Delta P^{\rm m}_{\mu e}$, the domain structure of
$\Delta A^{\rm CP}_{\mu e}(\delta)$ and $\Delta A^{\rm m}_{\mu e}$
can be understood through the ``solar" and ``atmospheric" magic
lines, and the interference phase condition, as suggested in
Ref.~\cite{Akhmedov:2008qt}. In the upper (lower) row of Fig.~4, we
give the numerical results of $\Delta A^{\rm m}_{\mu e}$ in NH (IH).
It is now evident that the most significant value appears in the
area of relatively short baselines and low energies, which are quite
relevant for the long-baseline experiments. The zoom-in plots of
this region are shown in the right column of Fig.~4, where the
ongoing and upcoming neutrino experiments have also been indicated
by solid diamonds and the energy ranges are represented by the peak
energies plus error bars. It is worthwhile to mention that $\Delta
A^{\rm m}_{\mu e}$ in Eq.~(12) is unchanged when we switch from NH
to IH through $\alpha \to -\alpha$, $\Delta \to -\Delta$, and $A \to
-A$. Hence, $\Delta A^{\rm m}_{\mu e}$ is insensitive to the
neutrino mass hierarchy.

\subsection{Optimal Experimental Setup}

Now, we discuss the optimal experimental setup to probe $\delta$.
The ongoing long-baseline experiments T2K ($E = 0.72\pm 0.27~{\rm
GeV}$, $L = 295~{\rm km}$) and NO$\nu$A ($E = 2.02\pm 0.43~{\rm
GeV}$, $L = 810~{\rm km}$), together with the proposed ones LBNE ($E
= 3.55\pm 1.38~{\rm GeV}$, $L = 1300~{\rm km}$) and
LAGUNA-LBNO\footnote{At the moment, the fate of this project is
unclear.} ($E = 5.05 \pm 1.65~{\rm GeV}$, $L = 2288~{\rm km}$), will
be considered for illustration. Except for T2K, all experiments are
intended to equally operate both in the neutrino and antineutrino
channels. Therefore, one can construct the CP asymmetry $A^{\rm
CP}_{\mu e}$ by measuring the neutrino and antineutrino
probabilities. In the right column of Fig.~4, we observe that these
experimental setups are lying on the ``first band" (i.e., $\Delta
A^{\rm m}_{\mu e} \sim 10~\%$), counted from top-right to
bottom-left.

To improve the experimental sensitivity, one can lower the neutrino
beam energy and locate the experiment on the ``second band" (i.e.,
$\Delta A^{\rm m}_{\mu e} \sim 15~\%$). However, this observation is
based on the probability level and we have to notice the energy
dependence of the neutrino flux and the cross section. In addition,
the detection efficiency and the background should be taken into
account. Therefore, a detailed simulation has to be performed in
order to make a final conclusion. Recently, it has been proposed
that the ESS proton beam can be adjusted to produce an intense
neutrino beam of energy around $0.4~{\rm GeV}$
\cite{Baussan:2012cw}. The simulation results indicate that with 8
years of data taking with an antineutrino beam and 2 years with a
neutrino beam up to $73~\%$ of the whole range of $\delta$ could be
covered at $3\sigma$ level at the optimal baseline of around
$500~{\rm km}$ \cite{Baussan:2012cw}. In the right column of Fig.~4,
the ESS proposal is shown as $E = 0.4 \pm 0.2 ~{\rm GeV}$ and $L =
500~{\rm km}$. Such an experimental setup happens to be on the
``second band", as we suggest. For preliminary performance of the
different experimental setups, see Refs.
\cite{Abe:2011ks,Abe:2011sj,NOvA,LBNE,LBNO,Baussan:2012cw}

It is worthwhile to mention that precision measurements of
atmospheric neutrinos in the PINGU detector at the IceCube
experiment on the South Pole may have good sensitivity to $\delta$
\cite{Akhmedov:2012ah}. A study of the PINGU detector and
accelerator-based neutrino beams also exists \cite{Tang:2011wn}. On
the other hand, there is no doubt that the future neutrino factories
are the best place to measure $\delta$ with high statistical
significance
\cite{Huber:2002mx,Huber:2003ak,Agarwalla:2010hk,Coloma:2012ji}.

\section{Summary}

We have made a complete survey of measures of leptonic CP violation
in neutrino oscillation experiments. Two new working observables
$\Delta A^{\rm CP}_{\alpha \beta}(\delta) \equiv A^{\rm CP}_{\alpha
\beta}(\delta) - A^{\rm CP}_{\alpha \beta}(0)$ and $\Delta A^{\rm
m}_{\alpha \beta} \equiv \max[A^{\rm CP}_{\alpha \beta}(\delta)] -
\min[A^{\rm CP}_{\alpha \beta}(\delta)]$ , where $A^{\rm CP}_{\mu
e}(\delta) \equiv P^{}_{\mu e}(\delta) - \bar{P}^{}_{\mu e}(\delta)$
is the CP asymmetry of oscillation probabilities, are suggested to
describe the intrinsic leptonic CP violation. Both analytical and
numerical calculations are performed to illustrate their main
features. The band structure of $\Delta A^{\rm m}_{\mu e}$ in the
baseline-energy plane can be implemented to optimize the
experimental setup. Furthermore, we have found that the current and
future long-baseline neutrino oscillation experiments are located on
the first band. On probability level, we observe that the decrease
of the neutrino beam energy in a proper way could improve the
experimental sensitivity.

Although the final verdict on the optimal experimental setup
requires a more sophisticated simulation, we expect our analysis to
be helpful in understanding the leptonic CP violation and useful in
probing the leptonic CP-violating phase in future long-baseline
experiments.

\acknowledgements

H.Z. would like to thank the G\"{o}ran Gustafsson Foundation for
financial support, and the KTH Royal Institute of Technology for
hospitality, where this work was initiated. The authors are indebted
to Evgeny Akhmedov and Mattias Blennow for useful discussions. This
work was supported by the Swedish Research Council
(Vetenskapsr{\aa}det), contract no. 621-2011-3985 (T.O.), the Max
Planck Society through the Strategic Innovation Fund in the project
MANITOP (H.Z.), and the G\"{o}ran Gustafsson Foundation (S.Z.).

\bibliography{bib}

\end{document}